\documentstyle[aps,prl]{revtex} 
\draft 
 
\def\lsim{\mathrel{\raise.3ex\hbox{$<$\kern-.75em\lower1ex\hbox{$\sim$}}}} 
\def\gsim{\mathrel{\raise.3ex\hbox{$>$\kern-.75em\lower1ex\hbox{$\sim$}}}}

\begin{document} 
 
\twocolumn[\hsize\textwidth\columnwidth\hsize\csname
@twocolumnfalse\endcsname
 
\title {MeV Dark Matter: Has It Been Detected?} 
\author{C\'eline Boehm$^1$, Dan Hooper$^1$, Joseph Silk$^{1,2}$, Michel Casse$^{2,3}$ and Jacques Paul$^{3}$} 
\address{
$^1$Denys Wilkinson Laboratory, Astrophysics Department, OX1 3RH  Oxford, England UK;
$^2$Institut d'Astrophysique de Paris;
$^3$CEA-Saclay, DSM/DAPNIA/Service d'Astrophysique, F-91191 Gif-sur-Yvette, 
France}
\date{\today} 
 
\maketitle 
 
\begin{abstract}

We discuss the possibility that the recent detection of 511 keV $\gamma$-rays from the galactic bulge, as observed by INTEGRAL, is a consequence of low mass ($\sim$MeV) particle dark matter annihilations. We discuss the type of halo profile favored by the observations as well as the size of the annihilation cross section needed to account for the signal. We find that such a scenario is consistent with the observed dark matter relic density and other constraints from astrophysics and particle physics.

\end{abstract}

]

\section{Introduction}

Recently, observations of a bright 511 keV $\gamma$-ray line from the
galactic bulge have been reported \cite{511}. This detection, made by the SPI
spectrometer on the INTEGRAL (INTErnational Gamma-Ray Astrophysics
Laboratory) satellite, suggests a spherically symmetric distribution of 511
keV $\gamma$-rays with a full width half maximum of $\sim 9^{\circ}$
($6^{\circ}-18^{\circ}$ at 2$\sigma$ confidence). The flux of the 511 keV
line emission from this region has been measured to be $9.9^{+4.7}_{-2.1}
\times 10^{-4}\,\rm{ph}\,\rm{cm}^{-2}\,\rm{s}^{-1}$ with a width of
 about 3 keV, in good agreement with previous measurements \cite{previous}.

The source of positrons in the galaxy
 is the subject of a great deal of debate. Some suggested sources include neutron stars or black holes \cite{compact}, radioactive nuclei from supernovae, novae, red giants or Wolf-Rayet stars \cite{stars}, cosmic ray interactions with the interstellar medium \cite{ism}, pulsars \cite{pulsars} and stellar flares. Despite this considerable list of potential sources, the origin of these positrons is still unknown today. Indeed, the most generous sources seem to be type Ia supernovae, however the fraction of positrons that escape from such an event, depending on internal mixing immediately after explosion and on the magnetic field strength and topology, is subject to debate \cite{debate}.

Massive Wolf-Rayet stars (hypernova), of the SN2003 type \cite{sn2003},
exploding in the galactic center could do the job \cite{casse}, but their
rate is unknown. Also, it is not known for certain that the positrons
injected could fill the whole galactic bulge, even if a bipolar galactic wind
is produced by star bursts \cite{starbursts}. Indeed, even if a ``galactic
positron fountain'' develops \cite{fountain}, the annihilation rate at high
altitude is too low, due to the small density of the wind, to explain the
extension of the 511 keV source \cite{pohl}.

Observations of the cosmic microwave background, the primordial abundances of
light elements and large scale structure have revealed that a great deal of
the mass of our universe consists of dark matter (DM)
\cite{darkmatterevidence}. Despite this large and growing body of evidence,
we are still ignorant of the nature of DM.

The most commonly discussed DM candidates are weakly interacting particles
with masses typically in the range of a few GeV to several TeV,
the most
popular example of this being the lightest supersymmetric particle
\cite{neutralino} made stable by the virtue of R-parity \cite{rparity}. This
particle, along with other fermionic DM particles, is not expected to be
significantly lighter than a few GeV if they are to provide the measured
density of DM in the universe \cite{lightfermions}. This constraint does not
strictly apply to all particle candidates, however. Light particles
($\sim$MeV-GeV), in some cases, may be able to satisfy relic density and
other constraints \cite{boehm2,boehm1}. In particular, light scalar DM
candidates are an interesting and viable possibility \cite{boehm1}.

In this Letter, we argue that light DM particles ($\sim$1-100 MeV)
annihilating into $e^+ e^-$ pairs in the galactic bulge may be the source of
the observed 511 keV emission line.

\section{Positron Propagation and Annihilation}

If DM particles of $\sim$1-100 MeV mass annihilate in the galactic bulge into
electron-positron pairs, the resulting positrons then travel with their
propagation being dominated by ionisation losses. This energy loss rate is
approximately given by \cite{longair}
\begin{equation}
\frac{dE}{dt} \sim 2 \times 10^{-9} \bigg(\frac{N_H}{10^5 \rm{m}^{-3}}\bigg)
(\ln \Gamma +6.6) \, \rm{eV/s}
\end{equation}
where $\Gamma$ is the positron's Lorentz factor and $N_H$ is the number
density of target atoms in the galactic bulge. This rate yields a stopping
distance of $\sim 10^{24}$ and $\sim 10^{26}$ cm for positrons of MeV and 100
MeV energy, respectively. For microgauss scale magnetic fields, the Larmor
radius is on the order of $10^{11}$ or $10^9$ cm for energies of 100 MeV and
1 MeV, respectively. Considering a simple random walk, the positrons are
roughly confined to a distance of $\sim \sqrt{R_{\rm{stop}} \times
R_{\rm{Larm}}}$, which is about a parsec, or less. Therefore the positrons
are easily confined to the galactic bulge as they lose their energy.

The cross section for the pair annihilation of a positron with an electron at
rest is simply, $\sigma_{\rm{pair ann}} \simeq (1\,\rm{bn})/\Gamma$. This
leads to a positron mean free path of
\begin{equation}
d(E) \sim \frac{1}{N_e \sigma_{\rm{pair ann}}} \sim \Gamma \,
\bigg(\frac{10^5 \, \rm{m}^{-3}}{N_e}\bigg) \, 10^{25} \, \rm{cm},
\end{equation}
where $N_e$ is the electron density. Thus, we see that a positron's stopping
distance is typically much shorter than its mean free path. Annihilations
are, therefore, expected to occur primarily for thermalized positrons,
producing a 511 keV line. For particles of $\sim$100 MeV mass, it may be
possible that a substantial fraction of the positrons could annihilate before
reaching non-relativistic energies. For this reason, lighter DM particles
($\sim$1-10 MeV) may be somewhat favored.
If the electron temperature in the galactic bulge is too low, positronium
formation may dominate, and results in a narrow line (25\% of the time) or 3-photon
continuum (75\% of the time), depending on the spin state of the
positronium.
 The data sugests that annihilations through positronium
dominate, resulting in a
narrow 511 keV line \cite{milne}.

\section{The Dark Matter Halo Profile}

Assuming an equal number of DM particles and anti-particles, the DM annihilation rate is proportional to the DM density squared. Independent of the particle mass or annihilation cross section, the distribution of 511 keV $\gamma$-rays from the galactic bulge can be used to constrain the DM halo profile of our galaxy.  

The halo profile is typically parameterized by
\begin{equation}
\rho(r) \propto \frac{1}{(r/a)^{\gamma}  [1+(r/a)^{\gamma}]^{(\beta-\gamma)/\alpha}} 
\end{equation}
where $\alpha$, $\beta$ and $\gamma$ are unitless parameters set by the choice of halo model and $a$ is the distance from the galactic center at which the power law breaks, also set by the halo model. For the inner kiloparsec or so, the region we are concerned with, $r$ is much smaller than $a$, and the profile description simplifies to
\begin{equation}
\rho(r) \propto \frac{1}{r^{\gamma}}. 
\end{equation}
The flux of annihilation products observed in a region of the sky is often described by the function $J(\Psi)$ which is proportional to the local DM density squared, integrated over the line of sight in the direction $\Psi$. For such a function to be meaningful, it should be averaged over the angular resolution of the experiment.

The distribution of events seen by INTEGRAL appears to fit reasonably well to a gaussian with full width half maximum of $\sim 9^{\circ}$, with a 2$\sigma$ confidence interval of $6^{\circ}-18^{\circ}$. In figure 1, we compared  distributions of $J(\Psi)$ averaged over the $2^{\circ}$ angular resolution of the SPI spectrometer, for a variety of halo profiles. Keeping in mind that the angular resolution used here is approximate and that there are large error bars for the width of the observed $\gamma$-ray distribution, this figure indicates that halo profiles with the modest slope of $\gamma \sim 0.4-0.8$ provide the best fit to the observed distribution. Such a value corresponds to a mild cusp less steep than the NFW distribution based on N-body simulations \cite{nfw}, but more steep than the model proposed by Kravtsov {\it et al.} which is based on observations of dwarf and low surface brightness galaxies \cite{krav}.

\begin{figure}[thb]
\vbox{\kern2.4in\includegraphics{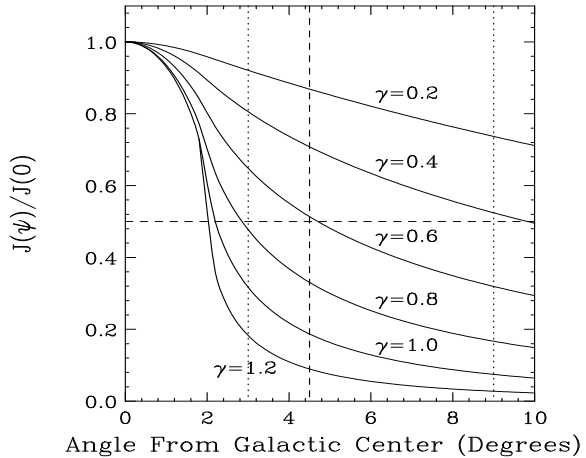}}

\caption{The angular distribution of $\gamma$-rays from DM annihilation averaged over the $2^{\circ}$ angular resolution of the SPI spectrometer on INTEGRAL for several halo profiles. SPI's observation indicates a full width half maximum of $9^{\circ}$ with a $6^{\circ}-18^{\circ}$ 2$\sigma$ confidence interval. Shown as vertical dashed and dotted lines are the central value and 2$\sigma$ limits of the angular widths found by SPI. To agree with this data, a halo model with $\gamma \sim$ 0.4-0.8 is favored.}
\end{figure}

\section{The Annihilation Cross Section}

If DM is made up of particles with masses below the muon and pion masses, low velocity annihilations can only produce electron-positron pairs, or perhaps photons or neutrinos. Assuming the former dominates, for each pair of DM particles which annihilate, one positron is produced, which eventually annihilates with an electron producing a pair of $\gamma$-rays. The flux of $\gamma$-rays produced in this scenario is given by
\begin{equation}
\Phi \cong 5.6 \, \bigg(\frac{\sigma v}{\rm{pb}}\bigg) \bigg(\frac{1\,\rm{MeV}}{m_{dm}}\bigg)^2 \overline{J}(\Delta \Omega) \Delta \Omega \, \rm{cm}^{-2} \rm{s}^{-1},
\end{equation}
where $m_{dm}$ is the mass of the DM particle, $\sigma v$ is the annihilation cross section multiplied by the relative velocity in units of $c$, $\Delta \Omega$ is the solid angle observed and $\overline{J}(\Delta \Omega)$ is the average of $J(\Psi)$ over that solid angle. We consider a solid angle of 0.02 sr, corresponding to a $9^{\circ}$ diameter circle. Again, we average over the $2^{\circ}$ angular resolution of the SPI spectrometer. For this solid angle and resolution, for an NFW halo profile ($\gamma$=1), we get $\overline{J}(0.02\,\rm{sr}) \simeq  187$. As we decrease the profile's slope, we get values of $\overline{J}(0.02\,\rm{sr})$ equal to 82.1, 37.3 and 17.1 for $\gamma$=0.8, 0.6 and 0.4, respectively.

Setting the calculated flux equal to the observed flux of $9.9 \times 10^{-4}\,\rm{ph}\,\rm{cm}^{-2} \,\rm{s}^{-1}$, we arrive at
\begin{equation}
\bigg(\frac{\sigma v}{\rm{pb}}\bigg) \bigg(\frac{1\,\rm{MeV}}{m_{dm}}\bigg)^2 \overline{J}(0.02\,\rm{sr}) \cong 0.003.
\end{equation}
Using the width of the observed distribution ($\overline{J} (0.02\,\rm{sr}) \sim $ 10-100), we can now estimate the required annihilation cross section. For a 1 MeV mass, $\sigma v$ must be $\sim 10^{-4}$ to $10^{-5}$ pb for annihilations in the galactic bulge. For 100 MeV mass, roughly 0.1 to 1 pb is required.

\section{Relic Density and Astrophysical Constraints}

\subsection{Relic Density}

We will now attempt to address the question of: is the range of annihilation cross sections found in section IV consistent with constraints from relic density and other measurements?

Assuming no coannihilations, the annihilation cross section required for a thermal relic is approximately  
\begin{equation}
\sigma v \simeq 0.2 \times \frac{x_F}{\sqrt{g_{\star}}} \bigg(\frac{\Omega_{\rm{dm}}h^2}{0.11}\bigg)^{-1} \, \rm{pb}
\end{equation}
where $x_F$=$m_{\rm{dm}}/T_F\simeq17.2+\ln(g/g_{\star})+\ln(m_{\rm{DM}}/\rm{GeV})+\ln\sqrt{x_F} \sim 12-19$ for particles in the MeV-GeV range, and $g$ and $g_{\star}$ are the number of internal and relativistic degrees of freedom, respectively.

\subsection{Astrophysical Constraints}

DM annihilations which occur during the recombination epoch can yield substantial fluxes of highly redshifted $\gamma$-rays, which may be possible to be observed today. It has been shown \cite{boehm2}, however, that for the class of particle we consider in this letter, such constraints are not violated.

Another possible constraint comes from our knowledge of nucleosynthesis. A DM candidate may yield too much $^4He$ photodissociation if very large S-wave cross sections are present \cite{nucleosynthesis}. However, we have checked this constraint and found that the cross sections we consider are consistent with all light element abundances.

\subsection{Assessment}

Combining the constraints from the INTEGRAL measurement and the relic density calculation, we can assess the prospect of a light DM particle being responsible for the 511 keV emission in the galactic bulge. The annihilation cross section needed for $\sim$MeV DM to provide the correct relic density is somewhat larger than what the INTEGRAL data appears to require. This apparent discrepancy can be solved, however, if $\sigma v \propto v^2$. We can write the cross section as $\sigma v \sim a + b\, v^2$, where $a$ and $b$ are some constants related to the S and P-wave terms. If $a \gg b$, then the annihilation cross section is roughly velocity independent and the constraints discussed above can only be reconciled in the heavy portion of the range we consider ($\sim 100\,$ MeV). If the cross section is S-wave suppressed ($b \gg a$), however, considerably lighter DM particles ($\sim 1\,$ MeV) are possible.

\section{Possible Models And Particle Physics Constraints}

Although DM is typically assumed to be heavier, 
masses below 1 GeV can satisfy eq.~7, 
despite the Lee-Weinberg limit, provided that their annihilation cross section is almost independent of the DM mass or if annihilations involve another light particle with weak couplings to ordinary matter. The DM candidate itself must also have suppressed couplings to the $Z$.  

The first possibility is easy to achieve with scalar DM particles annihilating
through the exchange of heavy particles ($>$100 GeV, to evade LEP constraints),
non-chirally coupled to ordinary fermions. In this case the $a$-term in the annihilation cross section is similar in size to the $b$-term and the annihilation cross section is only weakly dependent on velocity. This scenario can provide the observed rate of $\gamma$-rays and satisfy eq.~7 for masses $\sim$100 MeV, but not for considerably lighter particles, which require S-wave suppressed cross sections. 

The second possibility relies on the exchange of neutral and light particles. 
Assuming, for simplicity, that the DM is a scalar, this exchanged particle could be a new fermion with negligible or 
even null couplings to the $Z$. Alternatively, it could also be a new gauge boson from an extra $U(1)$, for example, called a $U$ boson. 
Light, neutral fermions provide a $v$-independent cross section but the final state of the annihilations (at tree level) would be neutrinos rather than $e^+ e^-$ or $\gamma$-rays. $U$ boson exchange, on the other hand, provides an annihilation cross section into $e^+ e^-$ pairs, which is S-wave suppressed ($b \gg a$).

The $U$ boson properties can be constrained through its additional contributions to the muon or electron anomalous magnetic moments. One finds that 
light DM particles can provide the correct relic density by selecting appropriate values of the coupling $U$-$\rm{dm}$-$\rm{dm}^*$ once the coupling $U$-$f$-$\bar{f}$ is determined from $g-2$ \cite{boehm1}.

From eq.~7, $\sigma v \sim \rm{pb}$ at the freeze-out epoch for 
$m_{dm}\sim$MeV ($x_F \sim 12$ and $g_{\star} \sim 10$). Given the decomposition: $\sigma v = a + b v^2$ ($a=0$ for a $U$), we find $b \sim 10 \,\rm{pb}$, thus $b v^2 \sim \rm{pb}$ 
at the freeze-out epoch (since $v^2 = T_{fo}/m_{dm} \sim 1/12$), and therefore, $\sigma v \sim 10^{-5} \rm{pb}$ in our galaxy. For $U$ boson exchange, therefore, only very light DM particles ($\sim$MeV) are viable DM candidates.

Both the $U$ boson and the DM particle would evade colliders limits despite their small 
masses, as studied in \cite{boehm1}. Indeed, the direct production of $U$ bosons ($e^+ e^- \rightarrow \gamma \, U$) is likely to remain unobserved due to the large 
background associated with the photon pair production. Also, 
the cross section  
$e^+ e^- \rightarrow \gamma \, \rm{dm} \, \rm{dm}^*$ appears 
to be $\sim 6 \times 10^{-6} \ \left(\frac{m_U}{\rm{MeV}}\right)^4 \ 
\left(\frac{m_{dm}}{\rm{MeV}}\right)^{-2} \ \left(\frac{E}{\rm{GeV}}\right)^{-2}\,\rm{pb}$ 
while the $e^+ e^- \rightarrow \gamma \, \nu \, \bar{\nu}$ cross section is 
$\sim 10^{-3} \ \left(\frac{E}{\rm{GeV}}\right)^{2} \, \rm{pb}$. 
Hence, the existence of light DM could have escaped low 
energy experiments searching for anomalous 
single photon events (especially if the cross section 
for $e^+ e^- \rightarrow \rm{dm} \, \rm{dm}^*$ is 
much smaller than the 
difference between the theoretical prediction of the cross section 
$e^+ e^- \rightarrow \nu \, \bar{\nu}$ for 3 neutrino families, 
and its measurement  at the $Z$ pole) \cite{hearty}. 

The two-body decay of the $U$ boson 
could provide a detectable signal, 
but given our couplings, the main channel is likely to be into two DM 
particles (in which case the decay would be dominated by 
invisible modes). On the other hand, a $U$-boson could, perhaps, be observed 
in nuclear transitions \cite{deboer}. 

We also note that, in addition to positronium formation, higher order annihilation diagrams may contribute to a continuum component of the $\gamma$-ray spectrum.

\section{Conclusions}

We find that the recent observations of the SPI spectrometer on INTEGRAL of 511 keV $\gamma$-rays from the galactic bulge are consistent with the annihilation of light ($\sim$1-100 MeV) dark matter particles. The angular width of the observed emission favors a mild cusp in the galactic halo profile with $\gamma \sim$ 0.4-0.8. 

For $\sim \,$MeV dark matter particles, an S-wave suppressed annihilation
cross section is required. Such a cross section would be predicted for
interactions in which a new, light gauge boson were exchanged. For heavier
dark matter particles ($\sim$100 MeV), a nearly constant cross section is
needed. For this case, the exchange of new, heavy fermions could be
responsible.

{\it Acknowledgements}: We would like to thank Bertrand Cordier for helpful discussions.

\end{document}